\documentstyle[12pt]{article}
 
\begin{document}
\vspace{-2.0cm}
\bigskip
\begin{center}
{\Large \bf Wigner's little group as a  generator of gauge transformations}
\vskip 0.5cm

{\bf R. Banerjee}\footnote{Invited talk at the Wigner Centennial Conference
held at Pecs, Hungary, July 8- 12, 2002. e-mail: rabin@bose.res.in; rabin@post.kek.jp}
\vskip 1.0 true cm
 
S. N. Bose National Centre for Basic Sciences \\
JD Block, Sector III, Salt Lake City, Calcutta -700 098, India. \\
 and\\
Institute of Particle and Nuclear Studies \\
High Energy Accelerator Research Organisation (KEK)\\
Tsukuba, Ibaraki 305-0801, Japan.
\end{center}
\bigskip
 
\centerline{\large \bf Abstract}

The role of Wigner's little group , as an abelian gauge generator
in different contexts, is studied.
\vskip 0.5cm
\begin{flushleft}
{\large Keywords} Gauge transformation, little group \\
{\large PACS No.(s)} 11.10.Kk, 11.15.-q, 04.20.Cv
\end{flushleft}

In his paper 
{\it On unitary representations of the inhomogeneous Lorentz group}, published in 1932 \cite{w},  Eugene Paul Wigner introduced the concept of little group and used it
to classify the elementary particles on the basis of their helicity/spin
quantum numbers. The topic of little group was a personal  favorite of Prof. 
Wigner. During the last years of his life,  he wrote a series of seven papers in collaboration with
Y. S. Kim elaborating the geometrical meaning  of the little group \cite{kim}. 
Meanwhile there were also studies regarding gauge generating aspects of the little group \cite{we,hk}. Recently, we have 
found  some more interesting and hitherto unknown  facets of the little group  
in generating the gauge transformations in various abelian gauge theories, including
topologically massive ones \cite{bc1,bcs1,bc2,sc, bcs2}. In this talk, I describe our work in this direction,  highlighting the major
results\footnote{
{\bf Notation:} Greek alphabets $\mu , \nu $ etc denote the space-time indices in
3+1 dimensions, letters $a, b, c$ etc stands for  2+1 dimensions and those from the middle of the 
alphabet, i.e. $i, j, k$ etc stands for 4+1 dimensions.}. 
 Further details can be found from the references given.

Wigner's little group is defined as the subgroup of homogeneous
 Lorentz group that leaves the
 energy-momentum vector of a particle invariant:
${W^\mu}_\nu k^\nu = k^\mu$.
In 3+1 dimensions,  the little group for a massive particle
is the  rotation group $SO(3)$. On the other hand, for a
massless  particle, the little group is the Euclidean group $E(2)$ which is a
semi-direct product of $SO(2)$ and $T(2)$ - the group of translations in the
2-dimensional plane. As is well known, both the
rotational groups $SO(3)$ and $SO(2)$ determine the classification of particles
on the basis of
their spin quantum numbers. One can
obtain the little group $E(2)$ as a particular limit of the rotation group
$SO(3)$ by Inonu-Wigner group contraction. However, while the significance of rotational groups was evident, the role of 
translational group remained a mystery for a long time. Weinberg and Han et.
al. \cite{we,hk} noticed that the translational group acts as a gauge generator in Maxwell
theory. Following Weinberg \cite{we},   one can find the  explicit representation of Wigner's
little group which leaves invariant the 4-momentum $k^\mu = (\omega, 0, 0, \omega)^T$ of a photon
of energy $ \omega $ moving in the $z$-direction,  to be
$W_4(p, q;  \phi) = W(p, q) R(\phi)$,
where
\begin{equation}
W(p, q) = W_4(p, q;  0) = \left( \begin{array}{cccc}
1+ \frac{p^2 + q^2 }{2} & p & q  & -\frac{p^2 + q^2 }{2} \\
p & 1 & 0 & -p \\
q & 0 & 1 & -q \\
\frac{p^2 + q^2 }{2} &  p & q  & 1 -\frac{p^2 + q^2 }{2}
\end{array}\right)
\label{1}
\end{equation}  
 is  a particular representation of the  translational subgroup  $T(2)$ of
the little group and $ R(\phi)$ represents a $SO(2)$ rotation about the $z$-axis. 
Note that the representations $W(p, q)$ and $R(\phi) $ of the
translation and rotation groups satisfy the relations
$W(p, q)W(\bar{p},\bar{q}) = W(p+\bar{p}, q+\bar{q})$ and $R(\phi)R(\bar{\phi}) = R(\phi + \bar{\phi})$.

We begin the discussion by showing that (1) acts as a gauge generator for the Maxwell theory.
The free Maxwell theory  has the equation of motion $\partial_\mu F^{\mu \nu} = 0$ which follows from the Lagrangian $
{\cal{L}} = -\frac{1}{4}F_{\mu \nu }F^{\mu \nu }$ where $F_{\mu \nu } =
\partial_\mu A_\nu - \partial_\nu A_\mu$.
The gauge field $A^\mu (x)$ for a single mode can be written as
$ A^{\mu}(x) = \varepsilon^{\mu}(k) e^{ik \cdot x}
$ suppressing the positive frequency part without any loss of generality. 
In terms of the
polarization vector $\varepsilon^{\mu}$, the gauge transformation $A_{\mu}(x) \rightarrow
A^{\prime}_{\mu} =  A_{\mu} + \partial_{\mu}f$ (where $f(x)$ is an arbitrary
scalar function) is expressed  as
$\varepsilon_{\mu}(k) \rightarrow \varepsilon^{\prime}_{\mu} =  \varepsilon_{\mu}(k) + if(k)k_{\mu}$
where $f(x)$ has been written as $f(x) = f(k)e^{ik \cdot x}$.
The equation of motion, in terms of the polarization vector,  will now be given
by
$k^2 \varepsilon^{\mu} - k^{\mu} k_{\nu} \varepsilon^{\nu} = 0
$. The massive excitation corresponding to $k^2 \neq 0$ leads to the solution
$\varepsilon^{\mu} \propto k^{\mu}$ which can  be gauged away. For
massless excitations  ($k^2 = 0$), the Lorentz condition $k_\mu \varepsilon^{\mu} =0$ follows immediately from the momentum space equation of motion.  Taking  
$k^{\mu} = (\omega, 0, 0, \omega)^T$, corresponding to a photon of energy 
$\omega$ propagating in the $z$
direction, and using the Lorentz condition, one can easily show that $\varepsilon^{\mu}(k)$ is gauge equivalent 
to the maximally reduced form\footnote{This method of obtaining the maximally reduced form of polarization vector/tensor of a theory is henceforth called the
`plane wave method'.}
$\varepsilon^{\mu}(k) =
(0, \varepsilon^1, \varepsilon^2, 0)^T$
displaying the two transverse degrees of freedom corresponding to
$\varepsilon^1$ and $\varepsilon^2$.
Under the action of the translational group $T(2)$ in (\ref{1}),  this polarization
vector transforms  as follows:
$\varepsilon^{\mu} \rightarrow \varepsilon^{\prime \mu} = {W^{\mu}}_{\nu}(p, q) \varepsilon^{\nu} = \varepsilon^{\mu} +
 \left( \frac{p\varepsilon^1 + q\varepsilon^2}{\omega}\right)k^{\mu}.$ 
It is now obvious that, this can be identified as a gauge transformation by
choosing $f(k)$ suitably. This shows that the translational subgroup $T(2)$
of Wigner's little group for massless particles acts as gauge generator in
free Maxwell theory \cite{we,hk}.  Similar conclusions hold also for Kalb-Ramond theory in 3+1 dimensions
\cite{bc1}. 

Next, consider
the $B\wedge F$ theory, which is a topologically massive gauge theory described by the Lagrangian
${\cal L} = -\frac{1}{4}F_{\mu \nu}F^{\mu \nu} + \frac{1}{12}H_{\mu \nu \lambda}H^{\mu \nu \lambda} - \frac{m}{6}\epsilon^{\mu \nu \lambda \rho} H_{\mu \nu \lambda} A_{\rho}
$ based on the  vector field $A_\mu$ , from which $F_{\mu \nu}$ is constructed and the antisymmetric tensor field
$B_{\mu\nu}$, which defines $H_{\mu \nu \lambda} =  \partial_\mu B_{\nu
\lambda} + \partial_\nu B_{\lambda
\mu} + \partial_\lambda B_{\mu
\nu}.$

It is
invariant under the combined gauge transformations $A_{\mu}(x) \rightarrow  A_{\mu} + \partial_{\mu}f$ and 
 $B_{\mu \nu} \rightarrow  B_{\mu \nu} + \partial_\mu F_\nu
(x)
- \partial_\nu F_\mu(x)$ where $f(x)$ and $ F_\mu(x)$ are arbitrary functions. 
 Employing the plane wave method
 one can show that the massless excitations of $B\wedge F$ theory are gauge 
artifacts and the maximally reduced form of the polarization vector and tensor corresponding to the massive physical
excitations are given by
\begin{equation}
\{\varepsilon^{\mu \nu}\} = \left( \begin{array}{cccc}
0 & 0 & 0 & 0 \\
0 & 0 & c & -b \\
0 & -c & 0 & a \\
0 & b & -a & 0
\end{array} \right), ~~~~~~~
\varepsilon^{\mu} = -i \left( \begin{array}{c}
0 \\
a \\
b \\
c
\end{array} \right)
\label{3}
\end{equation}
with the duality relation 
$\varepsilon^{\mu \nu} \varepsilon_{\mu} = 0$ between them \cite{bc1}. Correspondingly,
the momentum vector is $p^\mu = (m, 0, 0, 0)^T$.
Now, unlike in the case of Maxwell and KR theories, here the representation
$W(p, q)$ (\ref{1}) fails to be the generator of gauge transformations. 

However one can arrive at the representation of the  group that generates gauge transformations in $B\wedge F$ theory by considering the action of the matrix
\begin{equation}
D(p, q, r) = \left( \begin{array}{cccc}
1 & p & q & r \\
0 & 1 & 0 & 0 \\
0 & 0 & 1 & 0 \\
0 & 0 & 0 & 1
\end{array} \right)
\label{4}
\end{equation}
(where $p, q, r$ are real parameters)
on the polarization vector and tensor in (\ref{3}):
\begin{equation}\varepsilon^{\mu} \rightarrow \varepsilon^{\prime \mu} = {D^{\mu}}_{\nu}(p,q,r)\varepsilon^{\nu} = \varepsilon^{\mu} - \frac{i}{m}(p a +
q b + r c)p^{\mu}
\label{5}\end{equation}
\begin{equation}
\{\varepsilon_{\mu \nu}\} \rightarrow \{\varepsilon^{\prime}_{\mu \nu}\} = D(p,q,r) \{\varepsilon_{\mu \nu}\} D^T(p,q,r) = \{\varepsilon_{\mu \nu}\} + (\Delta \varepsilon)_ {\mu \nu}
\label{6}\end{equation}
where $(\Delta \varepsilon)_ {\mu \nu}=k_\mu F_\nu
(k)
- k_\nu F_\mu(k)$  and $ F_\mu(k)$ is an arbitrary function. 
It is obvious  
 that these reproduce the gauge transformations in the $B\wedge F$ theory \cite{bc1}. 

The group, of which $D(p, q, r)$ is a 
representation,  can be found by  noticing that 
$D(p,q,r) \times D(p', q', r') = D(p+p', q+q', r+r')$
which is the composition rule for the 3-dimensional translational group $T(3)$.
Moreover, the generators $P_1 = \frac{\partial D(p, 0, 0)}{\partial p}, 
P_2 = \frac{\partial D(0, q, 0)}{\partial q}$ and 
$P_3 = \frac{\partial D(0, 0, r)}{\partial r}$  of $D(p, q, r)$
are the same as those of $T(3)$ as can be seen from their Lie algebra
$[P_1, P_2] = [P_2, P_3] =  [P_3, P_1] =0.$ Thus the gauge generator for 
$B\wedge F$ theory is the representation (\ref{4}) of the translational group
$T(3)$. One may also notice that there are  three different embeddings of 
$T(2)$ within $T(3)$ and they preserve the 4-momentum of massless particles 
moving in the three different spatial directions.

Now we describe a method by which one can derive the gauge generating representation of
 translational group for topologically massive theories from the corresponding representation for ordinary gauge theories living in  one higher space-time dimensions. 
The starting point of this dimensional descent method is to note that one can interpret a massive particle 
 in $d$-dimensions as a massless particle in $d+1$-dimensions, 
with the mass being considered as the momentum component along the additional dimension \cite{blt}.
In its content, dimensional descent is related to the Inonu-Wigner group contraction. 

For example the  momentum and polarization vectors of Maxwell theory in 
5-dimensions are given by
$p^i = (\omega, 0,0,0, \omega )^T$ and $\varepsilon^i = (0, a_1, a_2, a_3, 0)^T$. 
  With the identification of $\omega$ with the mass $m$ and the deletion of the 
last columns (by the applying the projection operator $\cal P =$ diagonal$(1, 1,1,1,0)$), these vectors respectively becomes the rest frame momentum and 
polarization vectors of Proca model in 4-dimensions  which is equivalent to 
$B\wedge F$ theory \cite{bc1,bc2}. 
The 5-dimensional analogue $W_5(p,q,r)$ of (\ref{1}) generate gauge 
transformations in  Maxwell theory in that space-time dimension \cite{bc2}. That is,
$\varepsilon'^i = {W_5(p,q,r)^i}_j \varepsilon^j = \varepsilon^i + \frac{pa_1 +q a_2+r a_3}{\omega}p^i$.
Using the projection operator $\cal P $ one can project out
 the extra 5th dimension:
$\delta\bar{\varepsilon}^\mu = {\cal P}  \delta {\varepsilon}^i = \frac{pa_1 + qa_2 +ra_3}{\omega} p^\mu$.
This is precisely the gauge transformation of the polarization vector in  
$B\wedge F$ theory in 4-dimensions. From the form of this transformation,
one can readily read off the matrix representation of the group that generate the gauge 
transformation of the $B\wedge F$ theory in 4-dimensions to be (\ref{4}). 
Dimensional descent from 4 to 3 dimensions yields analogous results for
2+1 dimensional Maxwell-Chern-Simons theory \cite{bcs1,bc2}.

The above considerations are also valid for linearized gravity theories, both the usual as well as the topologically massive ones \cite{sc} .
The role of Wigner's little group in generating the star gauge invariance in 
noncommutative gauge theories  can be a possible extension of 
the present work. It is also of interest to construct and study the equivalent of
the little group in Ads space where there is a novel gauge transformation connected to
partially massless theories, as reported in \cite{dw1}.

\bigskip

{\bf Acknowledgements:} It is a pleasure to thank my collaborators, B.Chakraborty and T.Scaria.

\vfill\eject
\end{document}